\documentclass[%
reprint, 
 amsmath,amssymb,
 aps, prd,
]{revtex4-2}

\usepackage{graphicx}% Include figure files
\usepackage{dcolumn}% Align table columns on decimal point
\usepackage{bm}% bold math
\begin{document}

\preprint{APS/123-QED}

\title{\textbf{What Becomes of the Antiferromagnetic String \\
When Long-Range Néel Order Is Lost? — Exact Results
} 
}% 

\author{Yuri DiMaggio}

\altaffiliation{Formerly at:\\I. Institut für Theoretische Physik, Universität Hamburg, Hamburg, Germany \\ E-mail: yuri.dimaggio@gmx.de}

\date{\today}% It is always \today, today,

\begin{abstract}
Early string descriptions of doped antiferromagnets associate hole motion in a locally Néel-ordered background with a path-dependent antiferromagnetic string. We ask what becomes of this mechanism when long-range Néel order is lost while short-range antiferromagnetic correlations persist. In a singlet-bond representation, the underlying basis states do not assign an oriented \(A/B\) spin registration to each bond. Therefore, any surviving long-wavelength antiferromagnetic axis is naturally described as a director, \(\mathbf n\sim-\mathbf n\). Long-range Néel order can nevertheless provide a global \(S^2\) lift of this director field; when that long-range registration is lost, non-liftable configurations of the \(RP^2\) manifold become physically accessible, opening a topologically nontrivial \(\mathbb Z_2\) sector. Its elementary defects are half-vortices that globally minimize the \(RP^2\) \(\sigma\)-model energy within the nontrivial \(\mathbb Z_2\) sector. An \textit{a posteriori} energetic comparison further shows that the projective half-vortex texture is softer than the corresponding vector reference texture. A pair of half-vortices with opposite windings is logarithmically confined, \(V(R) = \kappa \ln\left(R/r_{\text{core}}\right)\), \(\kappa = \pi \rho_s /2\). The resulting confinement is carried not by a one-dimensional string but by an extended two-dimensional spin texture. Thus, the loss of long-range Néel order does not eliminate antiferromagnetic confinement but removes the global Néel lift.  This activates a distinct projective mechanism in which a local path-dependent string is replaced, at long wavelengths, by a global topological texture with logarithmic confinement.
\end{abstract}

\maketitle

\section{Introduction} \label{sec:intro}

Hole doping of an antiferromagnetic Mott insulator rapidly suppresses long-range Néel order. This effect is well established experimentally in the cuprates, where even a relatively small concentration of doped holes strongly reduces the antiferromagnetic ordering temperature and eventually destroys the long-range ordered state \cite{julien,birgenau,drachuck}. At the same time, substantial short-range antiferromagnetic correlations and spin-wave-like excitations persist well into the doped regime \cite{kastner,keimer,matsuda}. Thus, doping does not simply eliminate the antiferromagnetic background; rather, it converts a long-range ordered state into a locally correlated spin medium. It is this regime—without long-range vector Néel order but with surviving short-range antiferromagnetic correlations—that provides the starting point of the present work.

A natural microscopic mechanism of hole confinement in an antiferromagnetic background was proposed in the late 1980s by Kumar and Mohan \cite{mohan} and by Hirsch \cite{hirsch} and was subsequently developed in Refs. \cite{dimashko_1,dimashko_2}. In this picture, the motion of a hole through a Néel-ordered background leaves behind a string of displaced or disturbed spins. The magnetic energy grows approximately in proportion to the string length, providing a confining interaction between the objects attached to its ends. This idea has recently undergone a substantial revival in the geometric-string description of magnetic polarons. In particular, Grusdt and collaborators formulated spinon–holon and spinon–chargon bound states connected by fluctuating geometric strings and showed that such strings provide a useful microscopic description of magnetic polarons in the \(t\)-\(J_z\) and \(t\)-\(J\) models, including situations with strong short-range antiferromagnetic correlations but without perfect long-range order \cite{grusdt_2018,grusdt_2019}. 

The same geometric viewpoint has also been extended to two doped holes. A recent effective theory describes pairs of mobile holes as partons connected by a fluctuating confining string and analyses the internal structure and spectrum of such bound states \cite{grusdt_2023}. More recently, Schlömer, Bohrdt, and Grusdt proposed a hidden-antiferromagnetic state generated by fluctuating Néel domain walls, described by an emergent \(\mathbb{Z}_2\) lattice-gauge structure \cite{schlomer_2025}. In this framework open domain-wall ends were identified as half-vortex, or vison, excitations, and their possible BKT-type unbinding was anticipated. This program has now been developed further by De Paciani et al., who constructed an effective theory with open string ends carrying vorticities \(u=\pm 1/2\) and interacting logarithmically at long distances \cite{de_paciani_2026}.

A complementary topological approach was proposed earlier by Morinari, who argued that a doped hole may create a half-Skyrmion, or meron-like, spin texture in an \(S^2\) nonlinear \(\sigma\)-model. In that construction the hole carries a nonvanishing topological charge, and multi-half-Skyrmion configurations were considered even beyond the loss of antiferromagnetic long-range order \cite{morinari}. 

Thus, both string-based and topological descriptions suggest that a doped carrier can produce a spatially extended structure in the antiferromagnetic background. At shorter distances, where a locally coherent Néel background remains meaningful, microscopic geometric-string and domain-wall approaches provide a natural description. The present work addresses the complementary long-wavelength regime and, in particular, asks what becomes of the spin-order parameter when the relevant magnetic state remains in the singlet sector while long-range Néel registration is lost.

The quantum origin of the relevant distinction is already visible at the level of the spin Hilbert space. Site-spin states span the full Hilbert space, whereas the singlet sector may be spanned by valence-bond states built from two-spin singlets,
\begin{equation*}
    \left| \psi_{ij} \right\rangle = \frac{1}{\sqrt{2}} \left( a^{\dagger}_{i\uparrow} a^{\dagger}_{j\downarrow} - a^{\dagger}_{i\downarrow} a^{\dagger}_{j\uparrow} \right) \left| 0 \right\rangle.
\end{equation*}A singlet bond does not assign the two opposite spin projections to the two endpoints in an oriented way: the two classical Néel registrations are contained in the same quantum state. The singlet basis therefore does not itself define a director, but it does not encode that distinction as a local bond degree of freedom. If short-range antiferromagnetic correlations still support such an axis at long wavelengths, the corresponding order parameter is naturally unoriented,
\begin{equation*}
   \mathbf n\sim-\mathbf n,
\end{equation*}
and its target space is
\[
RP^2=S^2/\{\mathbf n\sim-\mathbf n\}.
\]This statement concerns the low-spin sector and does not imply that the full many-body ground state must always be a singlet: higher-spin states remain available in the full site Hilbert space. Rather, it identifies the natural order-parameter manifold once the relevant low-energy branch is singlet-like and no independent \(A/B\) spin registration is retained.

For a finite \(SU(2)\)-invariant antiferromagnet, the exact ground state may be a symmetry-restored singlet and can therefore be represented without choosing an oriented Néel axis. In the thermodynamic limit, however, long-range Néel order supplies an additional piece of information: a boundary condition or an infinitesimal staggered field can fix the \(A/B\) registration throughout the sample, thereby defining a global lift of the director field to an oriented staggered vector on \(S^2\). As long as such a global lift exists, the physically relevant configurations are globally orientable and the conventional \(S^2\) description is sufficient.

Once long-range Néel order is lost while short-range antiferromagnetic correlations persist, this global lift is no longer fixed at macroscopic distances. The local antiferromagnetic axis may remain well defined, but only as a director. The full projective configuration space then becomes physically accessible, including configurations that cannot be lifted globally to a single-valued \(S^2\) field. Locally the \(S^2\) and \(RP^2\) nonlinear \(\sigma\)-models have the same exchange energy and the same field equations; their difference is global:
\[
\pi_1(S^2)=0,\qquad \pi_1(RP^2)=Z_2.
\]
Thus, the loss of long-range Néel order does not create the director degree of freedom itself. It removes the global \(S^2\) lift and thereby activates the nontrivial \(Z_2\) sector of the projective theory.

The \(RP^2\) nonlinear \(\sigma\)-model is well known in systems with director order \cite{ivanov,ballesteros,fukugita,ueda}, and related  \(\mathbb{Z}_2\) defects have been extensively discussed in frustrated antiferromagnets, most notably on triangular lattices \cite{kawamura,dombre,southern,wintel,kawamura_2007,yamaguchi,kawamura_2010,kawamura_2011,aoyama,tomiyasu}. In those systems the nontrivial topology originates from geometrical frustration and from a different microscopic order-parameter structure. 

In the present work, by contrast, the projective description arises in an originally bipartite antiferromagnet rather than from geometrical frustration. The singlet-bond representation of the low-spin sector does not retain an oriented \(A/B\) registration as a local degree of freedom, while long-range Néel order can restore an effective global lift to an oriented \(S^2\) field. When this long-range registration is lost, the non-liftable sector of the \(RP^2\) manifold becomes physically accessible. Within this target space we identify the invariant sector
\[
RP^1=S^1/\{\mathbf n\sim-\mathbf n\}\subset RP^2,
\]
and study its elementary branch-point states—planar \(RP^1\) half-vortices.

We show exactly that the elementary planar \(RP^1\) half-vortex globally minimizes the \(RP^2\) \(\sigma\)-model energy within its fixed nontrivial \(\mathbb Z_2\) sector.  Our proof reveals that the elementary planar \(RP^1\) half-vortices are globally stable even in the isotropic model—\textbf{no easy-plane anisotropy is required}.

\begin{figure*}[t]
\centering
\includegraphics[width=0.61\textwidth, trim=0 0.1cm 0 0, clip]{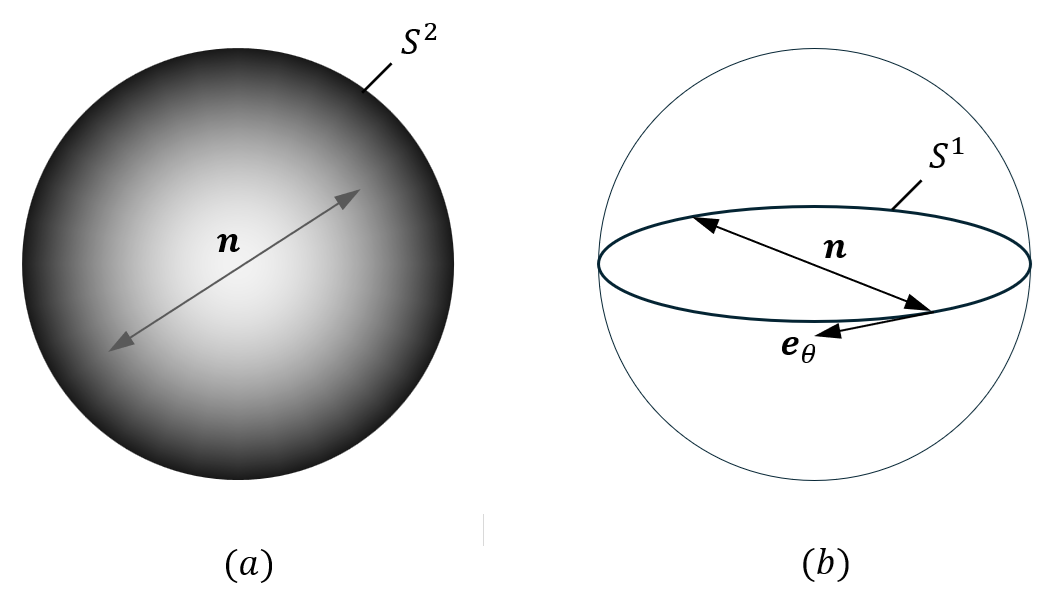}
\caption{a) The target space of the  \(RP^2 \;\sigma\)-model, \(RP^2=S^2/\{\bm{n} \sim  -\bm{n}\}\); b) the \(RP^1\) subspace of the full \(RP^2\) target space, \(RP^1=S^1/\{\bm{n} \sim  -\bm{n}\}\) - the great circle. The order-parameter field of  the full  \(RP^2\) target space is the Néel director \(\bm{n}\), which in the  great circle is specified by a single azimuthal angle  \(\theta\)  modulo \(\pi\).}
\end{figure*}

For a pair of elementary half-vortices with opposite windings, the resulting interaction is logarithmic,
\begin{equation*}
    V(R)=\kappa \ln\!\frac{R}{r_{\text{core}}}, \qquad \kappa=\frac{\pi \rho_s}{2}.
\end{equation*}Here, the interaction is not carried by a physical string. A cut may be introduced to connect the two branch points, but it is auxiliary and carries no line tension. The energy resides instead in the extended two-dimensional director texture. In this respect the present construction differs both from conventional microscopic string confinement and from recent theories in which half-vortices appear as endpoints of physical Néel domain walls. 

The paper is organized as follows. In Sec.~II we formulate the director \(RP^2\) nonlinear \(\sigma\)-model appropriate to the long-range-disordered but short-range-correlated regime. In Sec.~III we identify its invariant \(RP^1\) sector and construct the elementary branch-point, or half-vortex, configurations. Sec.~IV introduces the corresponding \(RP^1\) charge and establishes its conservation. Sec.~V proves the global stability of the elementary \(RP^1\) half-vortex within the full \(RP^2\) target space and the fixed nontrivial \(\mathbb Z_2\) sector. Sec.~VI derives the logarithmic confinement of opposite \(RP^1\)  charges.  Sec.~VII considers possible neutral and hole-carrying realizations of the confined states. Sec. VIII compares the resulting projective texture with microscopic string and vector reference configurations and clarifies its energetic advantage within the long-wavelength description. Sec.~IX summarizes the results and their implications.

\section{Spin order parameter: vector or director? } \label{sec: spin}

At length scales large compared with the lattice spacing, the magnetic sector of a square-lattice antiferromagnet is described by a nonlinear \(\sigma\)-model. For the exchange interaction \(J\), its static energy can be written as
\begin{equation}
    E[\bm{n}] = \frac{\rho_s}{2} \int (\nabla \bm{n})^2 \, d^2 r, \qquad \bm{n}^2 = 1.
\end{equation}where \(\rho_s\) is the spin stiffness. For the undoped spin-1/2 square-lattice Heisenberg antiferromagnet, \(\rho_s \approx 0.18J\) \cite{singh}; in the doped regime \(\rho_s\) will be understood as the effective stiffness of the remaining antiferromagnetic background. Minimization of energy (1) gives the Euler equation
\begin{equation}
  \nabla^{2} \bm{n} + (\nabla \bm{n})^{2} \bm{n}= 0.
\end{equation}The local form of energy (1), and consequently Eq. (2), is unchanged under the projective identification \(\bm{n}\sim - \bm{n}\): on any local lift of the director field, the replacement \(\bm{n}\rightarrow - \bm{n}\) leaves both expressions invariant. Thus, the local field equations do not distinguish between vector and director order parameters; the distinction is entirely global.

The microscopic origin of this distinction was discussed in the Introduction. In the singlet sector, the underlying bond states do not retain an oriented \(A/B\) spin registration as a local degree of freedom. This does not prevent an ordered antiferromagnet from admitting an oriented Néel-vector description. In the presence of long-range Néel order, a boundary condition or an infinitesimal staggered field fixes a consistent \(A/B\) registration throughout the sample and thereby defines a global lift of the local director field to a single-valued vector \(\bm n\in S^2\). In this case the conventional \(S^2\) description is sufficient.
Once long-range Néel order is lost while short-range antiferromagnetic correlations persist, this global lift is no longer fixed at macroscopic distances. A local antiferromagnetic axis may remain well defined, but only as an unoriented axis, \(\bm n\sim-\bm n\). The corresponding long-wavelength target space is therefore
\begin{equation}
  RP^2=S^2/\{\bm n\sim-\bm n\},
\end{equation} see Fig. 1. The passage from the globally liftable sector to the full \(RP^2\) configuration space does not modify the local exchange energy (1) or its Euler equation (2). What changes is the set of globally admissible configurations. Every globally oriented \(S^2\) configuration remains admissible after projection to \(RP^2\), but the projective theory additionally permits configurations that cannot be lifted globally to a single-valued \(S^2\) field. This distinction is topological:
\begin{equation}
 \pi_1(S^2)=0,
\qquad
\pi_1(RP^2)= \mathbb Z_2.
\end{equation}
Thus, long-range Néel order does not create the director degree of freedom; rather, it restricts the physically relevant configurations to those admitting a global \(S^2\) lift. When that long-range registration is lost, the non-liftable \(\mathbb Z_2\) sector of the \(RP^2\) manifold becomes physically accessible. In what follows we study the consequences of this enlarged configuration space.

\section{The \(RP^1\) invariant sector of \(RP^2\)}\label{sec:RP1}

The Euler equation (2) admits an invariant planar sector of the full \(RP^2\) target space. It is the projective great circle, see Fig. 1:
\begin{equation}
  RP^{1} = S^{1}/\{\bm{n}\sim -\bm{n}\}\subset RP^{2}.
\end{equation}States belonging to this subspace can be parameterized by a single azimuthal angle \(\theta\),
\begin{equation}
        \bm{n} = \bigl(\cos\theta,\sin\theta,0\bigr), \qquad \theta \equiv \theta + \pi.
\end{equation}Introducing the tangent vector   \(\bm{e}_{\theta} = \left(-\sin \theta,\ \cos \theta,\ 0\right)\), one finds
    \begin{equation}
        \nabla^{2}\bm{n} = -(\nabla \theta)^{2} \bm{n} + \bm{e}_{\theta}\,\nabla^{2}\theta.
    \end{equation}Substitution into Eq. (2) gives
\begin{equation}
        \nabla^{2}\theta = 0.
    \end{equation}Thus, the nonlinear \(RP^2\) field equation reduces exactly to the Laplace equation when restricted to the \(RP^1\) subspace. The corresponding static energy is
\begin{equation}
    E[\theta] = \frac{\rho_s}{2} \int (\nabla \theta)^2 \, d^2 r.
\end{equation}
The \(RP^1\) subspace is therefore not an approximate reduction but an exact invariant sector of the full theory.

The harmonic character of \(\theta\) allows the use of complex analysis. Let \(z=x+iy\) and introduce a locally analytic function \(W(z)=u(x,y)+i\theta (x,y)\), or, equivalently,
\begin{equation}
    f(z) = e^{W(z)} = e^{u + i\theta} = \lvert f(z) \rvert e^{i\theta}.
\end{equation}Then
\begin{equation}
    \theta(x,y)=\operatorname{Arg} f(z).
\end{equation}Here \(u\) and \(\theta\) are harmonic conjugates. The function \(f(z)\) is a holomorphic generating function for the planar \(RP^1\) texture in physical space. Because the physical director is unchanged under \(\theta \rightarrow \theta + \pi\), the transformations \(f(z) \rightarrow -f(z)\) describe the same physical state. Consequently, single-valuedness of the Néel director does not require \(f(z)\) itself to be single-valued; it is sufficient that \(f^2 (z)\) be single-valued. This permits the elementary branch-point solutions \(f(z)=(z-a)^{\pm 1/2}\), or
\begin{equation}
    \theta(z) = \pm \frac{1}{2}\operatorname{Arg}(z - a).
\end{equation}
Upon encircling the branch point \(z=a\), the angle changes by \(\pm \pi\), while the physical director returns to the same \(RP^1\) state. These are the elementary planar half-vortex configurations.

By linearity of Eq. (8), this construction extends immediately to an arbitrary set of branch points:
\begin{equation}
    \theta(z) = \frac{1}{2} \sum_i s_i \operatorname{Arg}(z - a_i), \quad s_i = \pm 1.
\end{equation}or, equivalently,
\begin{equation}
    f(z) = \prod_i \left( z - a_i \right)^{s_i/2}.
\end{equation}Because \(f(z)\) is generally multivalued, branch cuts may be introduced to specify a single branch. Their position and shape are arbitrary: continuous deformations of a cut that do not cross a branch point leave the physical director unchanged. The cut is therefore not a physical string and carries neither an independent energy nor a line tension. The energy resides in the extended \(RP^1\)  texture generated by the branch points.

\section{\(RP^1\) charge and its conservation}\label{sec:charge}

The branch-point configurations constructed in Sec.~III possess a natural topological charge. For any closed contour \(C\) that does not cross a defect core, we define the \(RP^1\) topological charge enclosed by \(C\) as
\begin{equation}
    w_{C} = \frac{1}{2\pi} \oint_{C} \nabla \theta \cdot d\ell
\end{equation}
Because the physical director is invariant under \(\theta \rightarrow \theta + \pi\), a closed path in physical space may change the lifted angle by an integer multiple of \(\pi\),
\begin{equation}
    \Delta_C \theta = m\pi,\quad m \in \mathbb{Z}.
\end{equation}
Hence
\begin{equation}
    w_C = \frac{m}{2}.
\end{equation}
The elementary branch-point states (12) correspond to \(m=\pm1\) and therefore carry charge
\begin{equation}
        w = \pm \frac{1}{2}.
    \end{equation}
For a general configuration (14) containing several branch points, the charge enclosed by \(C \) is additive,

\begin{equation}
    w_C = \sum_i w_i .
\end{equation}

Equivalently, using the holomorphic generating function \(f(z)\),
\begin{equation}
    w_C = \frac{1}{2\pi i}\oint_{C} \frac{f'(z)}{f(z)}\,dz
\end{equation} Thus, the exponents that appeared in the branch-point representation (14) acquire a direct topological meaning.

The \(RP^1\) charge \(w\) reflects the fundamental group \(\pi_{1}\!\left(RP^{1}\right) = \mathbb{Z}\) since \(RP^{1} \cong S^{1}\) . With the conventional \(2\pi\) normalization, the integer winding number of \(RP^1\) is represented by the half-integer quantity \(w=m/2\). Upon embedding \(RP^1\) into the full \(RP^2\) target space, however, only the parity of m remains topologically distinct,
\begin{equation}
  m \mapsto m \bmod 2
\end{equation}in accordance with the fundamental group of \(RP^2\), \(\pi_1 (RP^2)=\mathbb{Z}_2\). Therefore the two elementary states \(w=+1/2\)  and \(w=-1/2\)  represent opposite orientations of the \(RP^1\) winding but belong to the same nontrivial \(\mathbb{Z}_2\) class in \(RP^2\).

For smooth time evolution of the defects within the \(RP^1\) sector, we introduce, in the distributional sense, the space-time topological current
\begin{equation}
    j^\mu = \frac{1}{2\pi}\,\epsilon^{\mu\nu\lambda}\,\partial_\nu \partial_\lambda \theta,
\end{equation}
One then has identically
\begin{equation}
    \partial_\mu j^\mu = 0.
\end{equation}The temporal component is concentrated at the defect positions,
\begin{equation}
  j^{0}(r,t) = \sum_{i} w_{i}\,\delta^{(2)}\!\bigl(r - r_{i}(t)\bigr),
\end{equation}so that the total charge
\begin{equation}
    W = \int j^{0}\, d^{2}r = \sum_{i} w_{i}
\end{equation}is conserved unless charge crosses the boundary of the system. Creation or annihilation of defects is therefore possible only in combinations with vanishing net \(RP^1\) charge.

This conservation law refers to evolution within the invariant \(RP^1\) sector. If arbitrary excursions into the full \(RP^2\) target space are allowed, the integer \(RP^1\) winding is no longer a global topological invariant; only its parity, corresponding to the \(\mathbb{Z}_2\) class of Eq. (21), is protected. The physical relevance of the elementary \(w=\pm1/2\) defects therefore requires establishing their stability within the full \(RP^2\) target space, where only the \(\mathbb Z_2\) parity is topologically protected. This stronger question is addressed in the next section.

\section{Global stability of the elementary \(RP^1\) half-vortices}\label{sec:stability}

The invariance of the \(RP^1\) sector established above does not by itself guarantee its stability within the full \(RP^2\) target space. We now ask a stronger question: can an arbitrary finite deformation of an elementary \(RP^1\) half-vortex lower its energy while preserving the nontrivial \(\mathbb Z_2\) monodromy? The answer follows directly from the geometry of \(RP^2\).

Consider an annular region surrounding a single defect core,  \(r_{\rm core} < r < L\),
and introduce logarithmic radial coordinate
\begin{equation}
    \begin{aligned}
        &u=\ln\frac{r}{r_{\rm core}}, \qquad 0\leq u\leq U, \\
       &U=\ln\frac{L}{r_{\rm core}}, \qquad 0\leq\varphi<2\pi .
    \end{aligned}
\end{equation}In these coordinates the two-dimensional sigma-model energy takes the conformally invariant form
\begin{equation}
    E = \frac{\rho_s}{2} \int_{0}^{U} du \int_{0}^{2\pi} d\varphi \left[ \left| \partial_u \mathbf{n} \right|^2 + \left| \partial_\varphi \mathbf{n} \right|^2 \right].
\end{equation}

Throughout this section we consider continuous deformations at fixed defect core and fixed boundary conditions; such deformations cannot change the nontrivial \(\mathbb Z_2\) monodromy. For every fixed \(u\), the circle \(r=\mathrm{const}\) surrounds the same defect. Its image in the target space is therefore a closed loop \(\gamma_u\) belonging to the nontrivial homotopy class,
\begin{equation}
    [\gamma_u] \neq 0 \in \pi_1(RP^2) = \mathbb{Z}_2.
\end{equation}Lift this loop to the covering sphere \(S^2\). A nontrivial loop in \(RP^2\) lifts not to a closed path but to a path whose endpoints are antipodal,
\begin{equation}
    \widetilde{\bm n}(u,2\pi) = -\widetilde{\bm n}(u,0).
\end{equation}
Since the covering map \(S^2\rightarrow RP^2\) is locally isometric, the length of \(\gamma_u\) equals the length of its lift. Any path on \(S^2\) connecting antipodal points has length at least \(\pi\). Hence
\begin{equation}
    {\cal L}(u) = \int_0^{2\pi} \lvert \partial_\varphi \bm{n} \rvert \, d\varphi \geq \pi .
\end{equation}This is precisely what fails for an ordinary planar vortex embedded in \(S^2\): the corresponding target-space loop is contractible, so no positive lower bound on its length survives and the texture can escape out of the plane.

The Cauchy-Schwarz inequality then gives
\begin{equation}
    \int_{0}^{2\pi} \left|\partial_{\varphi} \bm{n}\right|^{2} \, d\varphi \geq \frac{{\cal L}^{2}(u)}{2\pi} \geq \frac{\pi}{2}.
\end{equation}

Substitution into Eq. (27), together with the non-negativity of the radial contribution, yields the global bound
\begin{equation}
    E \geq \frac{\rho_s}{2} \int_0^U du\,\frac{\pi}{2} = \frac{\pi \rho_s}{4} \ln \frac{L}{r_{\rm core}} \, .
\end{equation}Equality is possible only if both inequalities are saturated. This requires
\begin{equation}
    \partial_u \bm{n} = 0, \qquad \mathcal{L}(u) = \pi, \qquad \left| \partial_\varphi \bm{n} \right| = \frac{1}{2}.
\end{equation}Thus, at every radius the image of the physical circle must be a shortest path on \(S^2\) connecting antipodal points, traversed at constant speed. Such a path is one half of a great circle; after projection to \(RP^2\) it becomes a closed geodesic \(RP^1\subset RP^2\). Up to an arbitrary global rotation \({\cal R}\in SO(3)\), the minimizing configuration is therefore
\begin{equation}
    {\bm n}_{\rm min}(u,\varphi) = {\cal R} \left( \cos\frac{\varphi}{2},\, \sin\frac{\varphi}{2},\, 0 \right).
\end{equation}This is precisely the elementary \(RP^1\) half-vortex. Its energy saturates the bound,
\begin{equation}
    E_{\rm min} = \frac{\pi \rho_s}{4} \ln\frac{L}{r_{\rm core}} \, .
\end{equation}The result is stronger than linear stability. The elementary \(RP^1\) half-vortex is a global energy minimum among all continuous \(RP^2\) configurations carrying the same nontrivial  \(\mathbb Z_2\) monodromy on the annulus. A finite excursion of the director away from the \(RP^1\) great circle can only increase the energy: it either lengthens the corresponding target-space loop beyond the minimum value \(\pi\), introduces a radial dependence, or both. No easy-plane anisotropy is required.

The minimizing great circle is not unique, as expected for the isotropic model.  Different choices are related by global \(SO(3)\) rotations and reflect the rotational degeneracy of the isotropic model. The two orientations \(w=\pm1/2\) are related by a global rotation and therefore belong to the same \(SO(3)\) orbit of minimizers; in particular, they represent the same nontrivial \(\mathbb Z_2\) class and have identical energy

The argument is the \(RP^2\) analogue of the familiar statement that half-integer disclinations in nematics cannot escape into the third dimension \cite{cladis,ball}. In particular, the neutral two-defect geometry considered in Sec. VI, after compactification of the regular point at infinity, is conformally equivalent to an annulus.  The elementary \(RP^1\) half-vortex is therefore a global energy minimum at fixed defect core within the nontrivial \(\mathbb Z_2\) sector.

\section{\(RP^1\) confinement}\label{sec:confinement}

For a configuration of elementary defects with \(RP^1\) charges \(w_i=\pm 1/2\) at points \(a_i\), Eq. (13) can now be written as
\begin{equation}
    \theta(z) = \sum_i w_i \, \operatorname{Arg}(z - a_i), \qquad w_i = \pm \frac{1}{2}.
\end{equation}
Using
\begin{equation*}
    \nabla \operatorname{Arg}\!\left( z - a_i \right)
    = \hat{z} \times \nabla \ln \left| z - a_i \right|,
\end{equation*}and
\begin{equation*}
    \nabla^2 \ln \left| z - a_i \right|
    = 2\pi \delta^{(2)}\!\left( z - a_i \right),
\end{equation*}Eq. (9) gives
\begin{equation}
    E = \pi \rho_s W^2 \ln\!\frac{L}{r_{\mathrm{core}}} 
    - 2\pi \rho_s \sum_{i<j} w_i w_j \ln\! \frac{r_{ij}}{r_{\mathrm{core}}}
    + E_{\mathrm{core}}.
\end{equation}Here
\begin{equation}
    W = \sum_i w_i, \qquad r_{ij} = \lvert a_i - a_j \rvert,
\end{equation}\(L\) is the system size, \(r_{\mathrm{core}}\) is the short-distance cutoff, and \(E_{\mathrm{core}}\) contains the short-distance contributions not resolved by the continuum theory. The first term shows that a configuration with nonzero total \(RP^1\) charge has a logarithmically divergent energy in the thermodynamic limit \(L\rightarrow \infty\). Within the \(RP^1\) sector, finite-energy configurations in the thermodynamic limit therefore require vanishing total charge.

The simplest such state consists of two elementary defects with \(w_1=+ 1/2, w_2=-1/2\). For defects located at \(z=\pm a\), it is generated by
\begin{equation}
        f(z) = \sqrt{\frac{z-a}{z+a}}, \qquad
        \theta(z) = \frac{1}{2}\operatorname{Arg}\!\left(\frac{z-a}{z+a}\right).
\end{equation}
Their separation is \(R=2a\). Substitution into Eq. (37) gives the interaction energy
\begin{equation}
    V(R) = \kappa \ln\!\frac{R}{r_{\text{core}}}, \qquad \kappa = \frac{\pi \rho_s}{2}.
\end{equation}Remarkably, the same logarithmic coefficient is obtained independently in the recent effective theory of fluctuating Néel domain walls \cite{de_paciani_2026}: with \(M=2\pi\rho_s\) and endpoint vorticities \(u=\pm 1/2\), the pair interaction gives \(M/4=\pi\rho_s/2=\kappa\). 

The corresponding radial force is
\begin{equation}
    F_R = -\frac{dV}{dR} = -\frac{\kappa}{R}.
\end{equation}
Thus, two opposite elementary \(RP^1\) charges are logarithmically confined: their separation requires an energy that grows without bound, although more slowly than for a linear string potential.

This confinement should not be attributed to the branch cut introduced in representing the multivalued function (39). As discussed in Sec.~III, the cut is auxiliary and may be deformed without changing the physical state. There is no line tension associated with it. The energy is carried instead by the extended two-dimensional director texture through the energy density
\begin{equation}
    \varepsilon(r)=\frac{\rho_s}{2}\left(\nabla \theta\right)^2.
\end{equation}

\section{Physical realizations}\label{sec:realizations}

The analysis above determines the long-wavelength spin texture and the associated \(RP^1\) charges but does not specify the microscopic structure of the defect cores. In a doped \(t\)-\(J\) antiferromagnet, a branch point may therefore represent either an electrically neutral spin defect or a defect bound to a doped hole. This gives three elementary realizations of a confined pair of opposite \(RP^1\) charges, shown schematically in Fig. 2.

\begin{figure} [h!]
 \centering
 \includegraphics[scale=0.24]{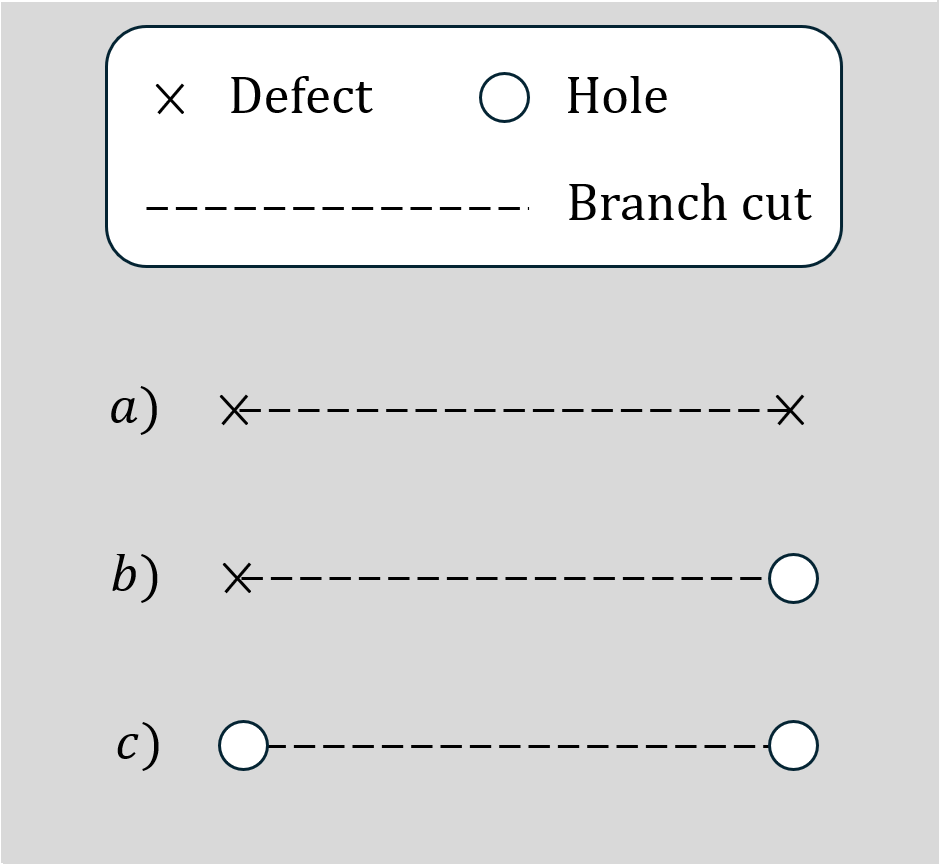}
\caption{Three basic physical realizations of a confined \(RP^1\) state. (a) A neutral defect–defect configuration, representing a \textit{topological spin excitation}. (b) A mixed defect–hole configuration, representing a \textit{topological spin-bag} of electric charge \(e\). (c) A hole–hole configuration, representing a \textit{topologically confined pair} of electric charge \(2e\). In all three cases the dashed line denotes an auxiliary branch cut of the generating function \(f(z)\), not a physical string. The two point-like carriers of opposite \(RP^1\) charges, located at the branch points, are logarithmically confined. }
\end{figure}

If both \(RP^1\) charges are carried by neutral spin defects, the resulting configuration has zero electric charge \(q=0\) and represents a purely topological spin excitation. If one of the two charges is associated with a hole of electric charge e, while the opposite charge remains electrically neutral, the confined state carries total charge \(q=e\). We shall refer to this dressed one-hole state as a topological spin-bag. Finally, if both opposite \(RP^1\) charges are associated with holes, the resulting state carries electric charge \(q=2e\) and represents a topologically confined pair of holes.

These three possibilities,
\begin{equation*}
    \begin{aligned}
        (0,+\tfrac{1}{2}) + (0,-\tfrac{1}{2}) &\longrightarrow q = 0, \\
        (0,+\tfrac{1}{2}) + (e,-\tfrac{1}{2}) &\longrightarrow q = e, \\
        (e,+\tfrac{1}{2}) + (e,-\tfrac{1}{2}) &\longrightarrow q = 2e,
    \end{aligned}
\end{equation*}may be summarized in Figure 2.

At the level of the long-wavelength \(RP^1\) description, the three states differ only in the conventional electric quantum numbers attached to the two carriers of opposite \(RP^1\) charge. Their long-distance spin interaction is governed by the same logarithmic confining potential (40).

The continuum theory fixes the common long-distance interaction of these states but not their core energies or their absolute binding energetics; these require an explicit treatment of the \(t\)-\(J\) model. A different question, however, can already be addressed at the continuum level: for a given configuration of defect endpoints, how do the representative long-wavelength magnetic structures compare energetically? We turn to this question next.

\section{Energetic comparison of magnetic structures}\label{sec:dynamics}

The continuum \(\sigma\) model does not determine absolute ground-state energies, nor can it establish whether a two-hole state is stable relative to two independently dressed holes; these questions require an explicit treatment of the microscopic \(t\)-\(J\) model. The role of the continuum theory is more limited. Once the projective sector is physically accessible, as argued in Sec. II, it allows one to compare the long-distance magnetic cost of representative textures compatible with a given configuration of defect endpoints.
This comparison should not be interpreted as an energetic selection of the \(RP^2\) target space itself. The director description follows from the singlet-bond structure of the low-spin sector together with the loss of a global \(S^2\) lift when long-range Néel registration disappears. The question addressed here is instead which magnetic texture is energetically favored within the resulting enlarged configuration space.

It is useful to compare three representative structures. They should not be regarded as three competing stable phases. The first two serve as reference configurations, whereas only the third is established here as a globally stable continuum solution.

The first reference is the microscopic antiferromagnetic string. In a locally Néel-ordered background, hole motion leaves behind a path-dependent sequence of disturbed bonds, with a magnetic energy growing approximately linearly with the endpoint separation \footnote{The lattice spacing is taken as unity.},
\begin{equation}
   E_{\rm str}(R)\sim JR .
\end{equation}This object is not a solution of the continuum \(\sigma\) model; it is a microscopic string ansatz whose cost is concentrated along a one-dimensional path.

A second reference is obtained when the magnetic disturbance is allowed to spread over the surrounding two-dimensional spin medium while retaining an oriented vector order parameter. A planar vector vortex then gives the softer logarithmic scaling
\begin{equation}
  E_v(R)\sim J\ln\frac{R}{r_{\rm core}} .
\end{equation}Such a vortex solves the planar field equation, but in the full isotropic \(S^2\) target space it is unstable: its target-space loop is contractible, and the texture can lower its energy by escaping out of the plane. It therefore serves only as an energetic reference for a delocalized vector texture.

The situation changes qualitatively in the non-liftable sector of the \(RP^2\) manifold. A loop around an elementary defect need traverse only a path from \(\bm n\) to \(-\bm n\) on the covering sphere, corresponding to a rotation by \(\pi\) rather than \(2\pi\). The elementary winding is therefore \(w=\pm1/2\). Since the \(\sigma\)-model energy is quadratic in gradients, the corresponding logarithmic exchange energy is reduced by a factor of four relative to the planar vector vortex,
\begin{equation}
  E_{\rm hv}(R)\sim\frac{J}{4}\ln\frac{R}{r_{\rm core}} .
\end{equation}More importantly, Sec. V established that this half-vortex is not merely a softer planar ansatz: it is the global minimum of the full isotropic \(RP^2\) model within the fixed nontrivial \(\mathbb Z_2\) sector.

The hierarchy may therefore be summarized schematically as
\begin{equation}
  JR
\;\longrightarrow\;
J\ln R
\;\longrightarrow\;
\frac{J}{4}\ln R .
\end{equation}It represents a sequence of progressively less constrained magnetic textures: a path-confined microscopic string, a spatially distributed but unstable vector vortex, and a globally stable projective half-vortex. The last step is qualitatively different from the first two because it becomes possible only when non-liftable \(RP^2\) configurations are physically accessible.

Recent fluctuating-domain-wall constructions \cite{schlomer_2025,de_paciani_2026} provide an intermediate realization of the same physical tendency. There, the descendant of the microscopic string survives as a fluctuating Néel domain wall, while its endpoints acquire vortex character. Such a state therefore combines features of the microscopic string and the extended topological texture. In the present long-wavelength projective description, the physical wall itself is no longer required: only the endpoint defects and the extended two-dimensional director texture remain.

The doped charges do not alter this long-distance comparison. At the continuum level, they determine which defect endpoints carry electric charge and how those endpoints move, whereas the long-distance magnetic dressing is governed by the spin sector. As shown in Sec. VII, the neutral, one-hole, and two-hole realizations differ in the electric quantum numbers attached to the endpoints but share the same logarithmic interaction. The microscopic \(t\)-\(J\) dynamics is still required to determine the core energies, kinetic contributions, and the actual stability of a two-hole state relative to two independently dressed holes.

The role of the energetic comparison can now be stated precisely. It does not select the director target space over the vector one; that distinction is fixed by the quantum structure of the low-spin sector and by the existence or absence of a global Néel lift. Rather, once the nontrivial projective sector is accessible, the continuum dynamics selects the elementary half-vortex as the energetically minimal long-wavelength dressing of the defect configuration.

\section{Conclusions and outlook}\label{sec:conclusions}

We have addressed the question of what becomes of antiferromagnetic string confinement once long-range Néel order is lost while short-range antiferromagnetic correlations persist. The central distinction is global rather than local. In a singlet-bond representation, the low-spin sector does not retain an oriented \(A/B\) spin registration as a local degree of freedom, so that any surviving long-wavelength antiferromagnetic axis is naturally described as a director, \(\bm n\sim-\bm n\). Long-range Néel order can nevertheless impose a global \(S^2\) lift of this director field by fixing the \(A/B\) registration throughout the sample. When this long-range registration is lost, the global lift is no longer fixed at macroscopic distances. Then, a non-liftable configurations of the \(RP^2\) manifold become physically accessible. Therefore, the local exchange functional is unchanged, but a topologically nontrivial \(\mathbb Z_2\) sector is activated.

Within this projective \(RP^2\) theory we have identified an exact invariant \(RP^1\) sector of the \(RP^2\) target space, \(RP^1 \subset RP^2\). Its elementary branch-point configurations are planar half-vortices with windings \(w=\pm 1/2\). Although the two signs represent opposite \(RP^1\) windings, both belong to the same nontrivial \(\mathbb{Z}_2\) class of the full \(RP^2\) target space. The corresponding \(RP^1\) charge is conserved under evolution within the invariant sector. Most importantly, the elementary planar \(RP^1\) half-vortices are global energy minima of the isotropic \(RP^2\) sigma model within the fixed nontrivial \(\mathbb Z_2\) sector. Any finite deformation away from the \(RP^1\) great circle either lengthens the corresponding nontrivial target-space loop beyond its minimum value \(\pi\), introduces radial gradients, or both. Thus, the half-vortex is stabilized by the topology and geometry of the isotropic \(RP^2\) model itself, \textbf{without invoking an easy-plane anisotropy}.

For a neutral pair of opposite elementary charges, the exact continuum interaction is
\begin{equation*}
    V(R) = \kappa \ln\!\frac{R}{r_{\text{core}}}, \qquad \kappa = \frac{\pi \rho_s}{2}.
\end{equation*}
This confinement is qualitatively different from a microscopic antiferromagnetic string. A branch cut may be introduced in representing the multivalued generating function, but its position is arbitrary and it carries no line tension. The energy resides instead in the extended two-dimensional director texture. The carrier of confinement has therefore changed from a local path-dependent object to a global topological configuration whose interaction depends only on the defect positions.

The present construction is closely connected with the recent geometric-string program. In the hidden-antiferromagnetic scenario of Schlömer, Bohrdt, and Grusdt \cite{schlomer_2025}, fluctuating physical Néel domain walls form a \(\mathbb{Z}_2\) string network, whose open ends are half-vortex excitations; the subsequent analysis by De Paciani et al. derives their logarithmic interaction and studies their BKT-type deconfinement \cite{de_paciani_2026}. The agreement of the logarithmic coefficient with Eq. (40) provides a nontrivial consistency check between the two descriptions.

Our construction nevertheless differs from Refs. \cite{schlomer_2025,de_paciani_2026} at a fundamental level. \textit{In the string-based approach}, within a locally coherent Néel background, the descendant of the microscopic antiferromagnetic string is a physical \(\pi\) domain wall, or stripe, terminated by two half-vortices. The confining object is therefore composite: two endpoint defects connected by a physical line defect with finite tension. \textit{In our projective description}, beyond the range of globally oriented Néel order, the physical stripe is absent. Only the two \(RP^1\) half-vortices and the extended two-dimensional director texture remain; the line connecting the branch points is merely an auxiliary cut and carries no tension. In this case the \(\mathbb Z_2\) structure follows directly from the topology \(\pi_1(RP^2)=\mathbb Z_2\) of the order-parameter space itself.

The two approaches are therefore complementary: the string construction describes the regime in which a locally coherent Néel background still supports a physical domain wall, whereas our projective construction describes the long-wavelength regime in which this physical wall is no longer defined.

\textbf{For the present construction, the decisive stabilizing ingredient is the nontrivial target-space \(\mathbb Z_2\) monodromy}. A loop encircling an elementary defect lifts to a path on \(S^2\) connecting antipodal points. Its length cannot be smaller than \(\pi\), and the minimum is attained only by a great semicircle, whose projection is an \(RP^1\) geodesic. The elementary \(RP^1\) half-vortex therefore does not merely represent a stationary configuration: it is the global energy minimum within its fixed nontrivial \(\mathbb Z_2\) sector.

The continuum theory fixes the universal topological sector and its long-distance confining interaction, but it also identifies the next necessary step. The defect cores must now be endowed with microscopic dynamics. This will require incorporating the mobile holes explicitly in the \(t\)-\(J\) model and restoring the underlying lattice discreteness. At that level one can determine the core energies, kinetic contributions, and the energetic competition among the neutral defect pair, the one-hole topological spin-bag, and the two-hole configuration introduced in Sec.~VII. In particular, it will then become possible to decide whether the topologically confined two-hole state is stable relative to two independently dressed holes.

The main conclusion is therefore twofold. First, the loss of long-range Néel order does not eliminate antiferromagnetic confinement; rather, by removing the global Néel lift it activates the non-liftable \(\mathbb Z_2\) sector of the projective order-parameter manifold. Second, once this sector is accessible, the long-wavelength magnetic energetics selects the elementary half-vortex as the the minimum-energy long-wavelength dressing within the nontrivial \(\mathbb Z_2\) sector. The resulting confinement is logarithmic and is carried not by a physical one-dimensional string but by an extended two-dimensional director texture. The exact \(RP^1\) solution identifies this projective long-wavelength sector, while its consistency with the geometric-string description establishes the complementarity of the microscopic string and continuum topological regimes. The next step is to incorporate the microscopic dynamics of the defect cores within the discrete \(t\)-\(J\) model.

\bibliography{references}% Produces the bibliography via BibTeX.

\end{document}